\documentstyle[12pt]{article}
\begin{document}

\hfill IF-UFRJ-22/94
\title{\Large\bf Symplectic quantization for reducible systems}

\author{J. Barcelos-Neto\thanks{\noindent Electronic mails: ift03001
@ ufrj and barcelos @ vms1.nce.ufrj.br} and M.B.D. Silva\\
Instituto de F\'{\i}sica\\
Universidade Federal do Rio de Janeiro\\
RJ 21945-970 - Caixa Postal 68528 - Brasil}

\maketitle
\abstract
We study an extension of the symplectic formalism in order to
quantize reducible systems. We show that a procedure like {\it
ghost-of-ghost} of the BFV method can be applied in terms of Lagrange
multipliers. We use the developed formalism to quantize the
antisymmetric Abelian gauge fields.

\vfill
\noindent PACS: 03.70.+k, 11.10.Ef, 11.15.-q
\vspace{1cm}
\newpage

\section{Introduction}

\bigskip
Systems where some of its constraints (first or second class) are not
independent are said to be reducible. The use of the Dirac
method \cite{Dirac} to quantize these systems requires the elimination
of dependent constraints which leads to a final set where just
independent ones are present. We mention that this is an opposit
procedure related to the quantization method due to Batalin, Fradkin
and Vilkovisky (BFV) \cite{BFV,H}.  There, the elimination of
dependent constraints (supposing that they are first-class) cannot be
done.  This would imply a lower number of ghost fields. In
consequence, it would be impossible to achieve any final covariant
result. The solution of this problem in the BFV method is the
introduction of extra ghosts, called {\it ghost-of-ghost}, one for
each relation among constraints.  An interesting example involving
this subject is the quantization of superparticle and superstrings,
where the covariant quantization is just achieved after the
introduction of an infinite tower of ghost-of-ghosts \cite{GSW}.

\medskip
Recently, Faddeev and Jackiw \cite{FJ} have remind us of the
possibility of using the symplectic formalism \cite{W} as an
alternative quantization method to the Dirac one. They use it to
quantize the chiral-boson \cite{CB}, which is not a constrained
system in the symplectic point of view, eventhough it is in the Dirac
case \cite{Gi}. Incidentally we mention that one of the interesting
aspects of the symplectic method is that primary constraints of the
Dirac formalism are not constraints in the symplectic case. That is
why chiral bosons have no constraints in the symplectic formalism.
Another example is the fermionic Dirac theory \cite{Go}. On the other
hand, when the Dirac constraint theory has second-class constraints,
these may also be constraints in the symplectic procedure (we call
them {\it true constraints}). In this case, the symplectic
formalism can also be applied since the phase space is properly
modified in order to consistently define the symplectic tensor
\cite{Eu1,Outros,Eu2}.

\medskip
The purpose of the present paper is to consider the symplectic
formalism for reducible systems. The elimination of dependent
constraints, as is done in the Dirac case, is not possible to be
applied here because this would imply in a reduction of the number of
Lagrange multiplies, which are important in the absorbtion of
superfluos degrees of freedom (The symplectic tensor can only be
defined after the elimination of these superfluous fields).  The
solution we propose for this problem is to parallel the BFV
formalism. We also introduce {\it Lagrange multipliers-of-Lagrange
multipliers} for each reducible relation. As an example, we quantize
the antisymmetric Abelian tensor gauge fields. The consistency of our
final results are verified in the comparation with the ones obtained
by using the Hamiltonian Dirac procedure~\cite{Dirac}.

\medskip
Our paper is organized as follows. In Sec. 2 we make a brief review
of the symplectic formalism including the case with (true)
constraints. At the end we address the way of dealing with the
reducible case. In Sec. 3 we apply the developed formalism to the
antisymetric Abelian gauge fields.

\section{Brief review of the simplectic formalism with
constraints}

Let us consider a dynamical system evolving in a phase space and
described by the canonical set of variables $(q_i,\,p_i)\hskip
.3cm(i=1,\dots,N)$. These satisfy the fundamental Poisson brackets
\footnote{We develop this section just using discrete
coordinates. The extension to the continuous case can be done in a
straightforward way.}

\begin{eqnarray}
\bigl\{q_i,\,q_j\bigr\}&=&0
=\bigl\{p_i,\,p_j\bigr\}\,,\nonumber\\
\bigl\{q_i,\,p_j\bigr\}&=&\delta_{ij}\,.\label{1}
\end{eqnarray}

\bigskip
\noindent Considering that the bracket of some quantity $A(q,p)$
with anything satisfies the relation

\begin{equation}
\bigl\{A(q,p),\cdots\bigr\}=\frac{\partial A}{\partial q_i}\,
\bigl\{q_i,\cdots\bigr\}+\frac{\partial A}{\partial p_i}\,
\bigl\{p_i,\cdots\bigr\}\,,
\end{equation}

\bigskip
\noindent and using the fundamental brackets (\ref{1}), one can write the
usual Poisson bracket relation involving two arbitrary quantities, say
$A(p,q)$ and $B(p,q)$, as

\begin{eqnarray}
\bigl\{A(q,p),\,B(p,q)\bigr\}&=&\frac{\partial A}{\partial q_i}\,
\bigl\{q_i,\,q_j\bigr\}\,\frac{\partial B}{\partial q_j}
+\frac{\partial A}{\partial q_i}\,\bigl\{q_i,\,p_j\bigr\}\,
\frac{\partial B}{\partial p_j}\nonumber\\
&+&\,\,\frac{\partial A}{\partial p_i}\,
\bigl\{p_i,\,q_j\bigr\}\,\frac{\partial B}{\partial q_j}
+\frac{\partial A}{\partial p_i}\,\bigl\{p_i,\,p_j\bigr\}\,
\frac{\partial B}{\partial p_j}\,,\nonumber\\
&=&\frac{\partial A}{\partial q_i}\,
\frac{\partial B}{\partial p_i}
-\frac{\partial A}{\partial p_i}\,
\frac{\partial B}{\partial q_i}\,.\label{1a}
\end{eqnarray}

\bigskip
In order to figure out the simplectic structure, we write
coordinates and momenta in just one set of $2N$ generalized
coordinates which we denote by
$y^\alpha\hskip.3cm(\alpha=1,\cdots,2N)$, in such a way that

\begin{eqnarray}
y^i&=&q_i\,,\nonumber\\
y^{N+i}&=&p_i\,.
\end{eqnarray}

\bigskip
\noindent Now, the fundamental Poisson brackets simply read

\begin{equation}
\label{2}
\bigl\{y^\alpha,\,y^\beta\bigr\}
=\epsilon^{\alpha\beta}\,,
\end{equation}

\bigskip
\noindent where the antisymmetric tensor $\epsilon^{\alpha\beta}$ is
given by the matrix

\begin{equation}
\Bigl(\epsilon^{\alpha\beta}\Bigr)
=\left(
\begin{array}{cc}
0&I\\-I&0
\end{array}
\right)\,,
\end{equation}

\bigskip
\noindent and $I$ is the $N\times N$ identity matrix. The Poisson
bracket involving two arbitrary quantities $A(y)$, $B(y)$ can also be
directly obtained in terms of the tensor $\epsilon^{\alpha\beta}$

\begin{eqnarray}
\bigl\{A(y),\,B(y)\bigr\}
&=&\partial_\alpha A\,\bigl\{y^\alpha,\,y^\beta\bigr\}\,
\partial_\beta B\,,\nonumber\\
&=&\epsilon^{\alpha\beta}\,\partial_\alpha A\,
\partial_\beta B\,,
\end{eqnarray}

\bigskip
\noindent where $\partial_\alpha=\partial/\partial y^\alpha$. The
quantity $\epsilon^{\alpha\beta}$ guarantees the usual antisymmetry
of the Poisson bracket relations (for the bosonic case).

\medskip
By the observation of (\ref{2}) one might infer that the general form
of the brackets in the case with constraints is

\begin{equation}
\label{3}
\bigl\{y^\alpha,\,y^\beta\bigr\}
=f^{\alpha\beta}(y)\,,
\end{equation}

\bigskip
\noindent where $f^{\alpha\beta}$ is an antisymmetric tensor and that
must be nonsingular (notice that this last property is also verified
by $\epsilon^{\alpha\beta}$). Its inverse is the simplectic tensor
related to the constrained system. We mention that the simplectic
tensor can be used as a metric (simplectic metric) that raises and
lowers indices in the simplectic manifold~\footnote{Sympletic
manifolds whose metric is $\epsilon^{\alpha\beta}$ are related to
systems without constraints.  This is the case, for example, of the
self-dual fields \cite{FJ,CB}.}.

\medskip
Here, one can point out what are the general fundamentals of Dirac
and simplectic methods. The first one is developed by looking at the
left-hand side of expression (\ref{3}), that is to say, it tries to
generalize the Poisson brackets by including the constraints, until a
final form is reached where constraints can be taken as strong
relations  \cite{Dirac}. In the case of the simplectic formalism,
constraints (which are usually in a small number than in the Dirac
method) are used to make deformation in the geometrical structures in
order that the simplectic tensor can be consistently defined.

\medskip
The FJ formalism deals with first-order Lagrangians. It is opportune
to mention that this is not a serious restriction because all systems
we know, described by quadratical Lagrangians, can always be set in
the first-order formulation. This is achieved by extending the
configuration space with the introduction of auxiliary fields. These
are usually the momenta, but this is not necessarily so
 \cite{Eu1,Outros}. We mention that systems with higher
derivatives can be described in this same way~\cite{Eu2}.

\medskip
Let us consider a system described by a first-order Lagrangian like

\begin{equation}
L=a_\alpha(y)\,\dot y^\alpha-V(y)\,,
\end{equation}

\bigskip
\noindent where $y^\alpha$ is a set of $2N$ coordinates. $y^{i+N}$
can be the momenta or other auxiliary quantities introduced in order
to render the Lagrangian the first-order condition.  From the
expression above, the Euler-Lagrange equation of motion reads

\begin{equation}
\label{4}
f_{\alpha\beta}\,\dot y^\beta
=\partial_\alpha V\,,
\end{equation}

\bigskip
\noindent where

\begin{equation}
f_{\alpha\beta}=\partial_\alpha a_\beta
-\partial_\beta a_\alpha\,.
\end{equation}

\bigskip
\noindent If $\det\,(f_{\alpha\beta})\not=0$, one can solve (\ref{4})
for the velocities $\dot y^\alpha$, i.e.

\begin{equation}
\dot y^\alpha=f^{\alpha\beta}\,\partial_\beta V\,,
\end{equation}

\bigskip
\noindent where $f^{\alpha\beta}$ is the inverse of
$f_{\alpha\beta}$. This is the simplectic tensor reported earlier and,
in fact, one can show that $f^{\alpha\beta}$ are the Dirac bracket
between the coordinates $y^\alpha$, $y^\beta$  \cite{Go}.

\medskip
An interesting and instructive point occurs when the quantity
$f_{\alpha\beta}$ is singular. In this case one cannot identify it as
the simplectic tensor and, consequently, the brackets structure of the
theory cannot be consistently defined either. This means that the
system, even in the FJ approach, has constraints (true constraints).
One way to solve this problem is to follow the standard Dirac
formalism. However, this can also be achieved by working in a
geometric manner. In this case, we use the constraints to
conveniently deform the singular tensor quantity in order to obtain
another tensor that may be nonsingular.  If this occurs, this new
quantity can be identified as the simplectic tensor of the theory. Let
us briefly review the developments of the simplectic method when there
are true constraints involved  \cite{Eu1}.

\medskip
Let us denote the above mentioned singular quantity by
$f^{(0)}_{\alpha\beta}$, and suppose that it has, say, $M$
$(M<2N)$ zero modes $v^{(0)}_m$, $m=1,\cdots,M$, i.e.

\begin{equation}
\label{5}
f^{(0)}_{\alpha\beta}\,v^{(0)\beta}_m=0\,.
\end{equation}

\bigskip
\noindent The combination of (\ref{4}) and (\ref{5}) gives

\begin{equation}
\label{6}
\tilde v_m^{(0)\alpha}\,\partial_\alpha V^{(0)}=0\,.
\end{equation}

\bigskip
\noindent This may be a constraint. Let us suppose that this actually
occurs (we shall discuss the opposite case soon). Usually,
constraints are introduced in the potential part of the Lagrangian by
means of Lagrange multipliers. Here, in order to get a deformation in
the tensor $f^{(0)}_{\alpha\beta}$ we introduce them into the kinetic
part instead. This is done by taking the time-derivative of the
constraint and making use of some Lagrange multiplier
\footnote{It is well-known that constraints satisfy the
consistency condition of not evolving in time, that is to say, if
$\Omega$ is a constraint we have that $\dot\Omega$ is also a constraint.
Another point is that one could, instead, take the time derivative of
the Lagrange multiplier.}.

\medskip
These Lagrange multipliers, which we denote by $\lambda^{(0)}_m$,
enlarge the configuration space of the theory. This permit us to
identify  new vectors $a_\alpha^{(1)}$ and $a_m^{(1)}$ as

\begin{eqnarray}
a_\alpha^{(1)}&=&a_\alpha^{(0)}+\lambda_m^{(0)}\,
\partial_\alpha\Omega_m^{(0)}\,,\nonumber\\
a_m^{(1)}&=&0\,,
\end{eqnarray}

\noindent where $\Omega_m^{(0)}$ are the constraints obtained from
(\ref{6}). In consequence, one can now introduce the tensor quantities

\begin{eqnarray}
f_{\alpha\beta}^{(1)}&=&\partial_\alpha\,a_\beta^{(1)}
-\partial_\beta\,a_\alpha^{(1)}\,,\nonumber\\
f_{\alpha m}^{(1)}&=&\partial_\alpha\,a_m^{(1)}
-\partial_m\,a_\alpha^{(1)}=-\partial_m\,a_\alpha^{(1)}\,,\nonumber\\
f_{mn}^{(1)}&=&\partial_m\,a_n^{(1)}
-\partial_n\,a_m^{(1)}=0\,.
\end{eqnarray}

\noindent Here $\partial_m=\partial/\partial\lambda^m$. If $\det
f^{(1)}\not=0$, where $f^{(1)}$ is a matrix which also
involves the Lagrange multipliers, then we have succeeded in
eliminating the constraints. If not, one should repeat the procedure
above as many times as necessary.

\medskip
It may also occur that we arrive at a point where we still obtain a
singular matrix and the corresponding zero modes do not lead to any
new constraint. This is the case, for example, of gauge theories. At
this point, if we want to define the simplectic tensor, we have to
introduce some gauge condition. For details, see reference  \cite{Eu1}

\medskip
In order to have a clearer idea of the problem to be circumvented in
the case of reducible systems, we emphasize the role played by the
Lagrange multipliers in the symplectic formalism. They absorb some
superfluous degrees of freedom of the theory. This and the
deformation of the symplectic structure make possible the definiton
of the symplectic tensor, which is the final goal of the formalism.
Its inverse makes the bridge to commutators of the quantum sector.
So, when constraints are not independent and we eliminate some of
them as in the Dirac procedure, we have also a lower number of
Lagrange multipliers. This implies that some superfluous degrees of
freedom cannot be eliminated and, consequently, we are not able to
identify the symplectic tensor.

\medskip
We solve this  problem by paralleling the BFV procedure. We introduce
new Lagrange multipliers ({\it Lagrange multiplier-of-Lagrange
multiplier}) for each relation among the constraints and manipulate
these as in the previous case.

\section {An example involving reducible contraints}

Let us consider the case of abelian tensor gauge fields described by
the Lagrangian

\begin{equation}
\label{3.1}
{\cal L}=-\,\frac{1}{6}\,F_{\mu\nu\rho}\,F^{\mu\nu\rho}\,,
\end{equation}

\bigskip
\noindent where $F_{\mu\nu\rho}$ is a totally antisymmetric tensor
which can be written in terms of potential fields $A_{\mu\nu}$ (also
antisymmetric) by the relation

\begin{equation}
F_{\mu\nu\rho}=\partial_\mu A_{\nu\rho}
+\partial_\rho A_{\mu\nu}+\partial_\nu A_{\rho\mu}\,.
\end{equation}

\bigskip
\noindent This tensor is invariant (and consequently the Lagrangian
above) under the gauge transformations

\begin{equation}
A_{\mu\nu} \longrightarrow A_{\mu\nu}
+\partial_\mu\Lambda_\nu-\partial_\nu\Lambda_\mu\,.
\end{equation}

\bigskip
\noindent The reducible aspect of this theory can be envisaged from
the fact that if we choose the gauge parameter $\Lambda_\mu$ as the
derivative os some scalar quantity we will obtain that $A_{\mu\nu}$
do not change under the gauge transformation.

\medskip
This theory was already discussed from the Dirac  \cite{Kaul} and the
BFV  \cite{H} points of view. In order to use the symplectic method,
it is necessary to write the Lagrangian (\ref{3.1}) in the first
order notation.  First we rewrite it as

\begin{eqnarray}
{\cal L}=-\,{1\over2}\,\dot A_{ij}\dot A^{ij}
-2\partial_i A_{j0}\dot A^{ij}
&-&\partial_iA_{0j}\partial^iA^{0j}
+\partial_iA_{0j}\partial^jA^{0i}\nonumber\\
&-&{1\over2}\,\partial_iA_{jk}\partial^iA^{jk}
+\partial_iA_{jk}\partial^jA^{ik}\,.
\label{3.4}
\end{eqnarray}

\bigskip
\noindent It is opportune to mention that the symplectic method is
essentialy a noncovariant one. A first order version of this
Lagrangian reads

\begin{equation}
\label{3.5}
{\cal L}^{(0)}=\pi_{ij}\dot A^{ij}
+{1\over2}\,\pi_{ij}\pi^{ij}
+2\partial_iA_{j0}\pi^{ij}
-{1\over2}\,\partial_iA_{jk}\partial^iA^{jk}
+\partial_iA_{jk}\partial^jA^{ik}\,,
\end{equation}

\bigskip
\noindent where $\pi^{ij}$ is the auxiliary field (here it is the
momentum conjugate to $A_{ij}$). Its equation is a constraint
relation whose equation of motion leads to the initial
Lagrangian~(\ref{3.4}).

\medskip
{}From expression (\ref{3.5}) one identifies the quantitities

\begin{eqnarray}
a^{(0)A}_{\phantom{(0)}{ij}}&=&\pi_{ij}\,,\nonumber\\
a^{(0)A}_{\phantom{(0)}0j}&=&0\,,\nonumber\\
a^{(0)\pi}_{\phantom{(0)}ij}&=&0\,,
\end{eqnarray}

\bigskip
\noindent and obtains the tensors

\begin{eqnarray}
f^{(0)A\pi}_{\phantom{(0)}ijkl}(\vec x,\vec y\,)&=&
\frac{\delta a^{(0)\pi}_{\phantom{(0)}kl}(\vec y\,)}
{\delta A_{ij}(\vec x\,)}
-\frac{\delta a^{(0)A}_{\phantom{(0)}ij}(\vec x\,)}
{\delta\pi_{kl}(\vec y\,)}\,,\nonumber\\
&=&-\,\frac{1}{2}\,\Bigl(\delta_{ik}\delta_{jl}
-\delta_{il}\delta_{jk}\Bigr)\,
\delta^{(3)}(\vec x-\vec y\,)\,.
\end{eqnarray}

\bigskip
\noindent The remaining terms are zero. We then construct the matrix

\begin{equation}
f^{(0)}=\left(
\begin{array}{ccc}
0&0&0\\
0&0&-{1\over2}\bigl(\delta_{ik}\delta_{jl}
-\delta_{il}\delta_{jk}\bigr)\\
0&{1\over2}\bigl(\delta_{ik}\delta_{jl}
-\delta_{il}\delta_{jk}\bigr)&0
\end{array}
\right)
\,\delta^{(3)}(\vec x-\vec y\,)\,.
\end{equation}

\bigskip
\noindent We easily see that it is singular. Let us consider that a
zero mode has the general form $\tilde v^{(0)}=(v_k,\,u_{kl},\,
\omega_{kl})$. We thus get the equations

\begin{eqnarray}
\bigl(\delta_{ik}\delta_{jl}
-\delta_{il}\delta_{jk}\bigr)\,\omega_{kl}
&=0\Rightarrow\omega_{kl}=0\,,\\
\bigl(\delta_{ik}\delta_{jl}
-\delta_{il}\delta_{jk}\bigr)\,u_{kl}
&=0\Rightarrow u_{kl}=0\,,
\end{eqnarray}

\bigskip
\noindent and the quantities $v_k$ remain indeterminated.
Consequently, from the equation

\begin{equation}
\int d^3\vec x\,v_k(\vec x\,)\,
\frac{\delta}{\delta A_{0k}(\vec x\,)}\,
\int d^3\vec y\,V^{(0)}=0\,,
\end{equation}

\bigskip
\noindent that is the corresponding in the continuous of the
expression (\ref{6}) with $V^{(0)}$ given by

\begin{equation}
V^{(0)}=-{1\over2}\,\pi_{ij}\pi^{ij}
-2\partial_iA_{j0}\pi^{ij}
+{1\over2}\,\partial_iA_{jk}\partial^iA{jk}
-\partial_iA_{jk}\partial^jA^{ik}\,,
\end{equation}

\bigskip
\noindent one has

\begin{equation}
\int d^3\vec x\,v_k(\vec x\,)\,\partial_i\pi^{ik}=0\,.
\end{equation}

\bigskip
\noindent Since $v_k$ is a generic function of $\vec x$, we obtain
the constraints

\begin{equation}
\label{3.13}
\partial_i\pi^{ij}=0\,,
\end{equation}

\bigskip
\noindent which plays the role of Gauss' laws in this extension of
the electromagnetic theory. In consequence of the property of
reducibility we have that the constraints above are not independent.
Notice that $\partial_i\partial_j\,\pi^{ij}=0$.

\medskip
Now, what we have to do is to introduce the constraints (\ref{3.13})
into the kinectic part of the Lagrangian by means of Lagrange
multipliers.  Since these constraints are not
independent, we restric the Lagrange multipliers by means of some
convenient relation. In virtue of the similarity with the BFV case,
we restrict the Lagrange multipliers as

\begin{equation}
\label{3.14}
\partial^i\lambda_i=0
\end{equation}

\bigskip
\noindent and use another Lagrange multiplier ({\it Lagrange
multiplier-of-Lagrange multiplier}) to also introduce it into the
kinectic part of the Lagrangian. We then have

\begin{equation}
{\cal L}^{(1)}=\pi_{ij}\dot A^{ij}
-\dot\pi_{ij}\partial^i\lambda^j
-\dot\lambda^i\partial_i\eta-V^{(1)}\,,
\end{equation}

\bigskip\noindent
where

\begin{equation}
V^{(1)}=-{1\over2}\,\pi_{ij}\pi^{ij}
+{1\over2}\,\partial_iA_{jk}\partial^iA^{jk}
-\partial_iA_{jk}\partial^jA^{ik}\,,
\end{equation}

\bigskip\noindent
where the fields $A_{0j}$ were absorbed in $\dot\lambda_j$. Now, the
new coefficients are

\begin{eqnarray}
a^{(1)A}_{\phantom{(1)}ij}&=&\pi_{ij}\,,\nonumber\\
a^{(1)\pi}_{\phantom{(1)}ij}&=&\frac{1}{2}\,
\bigl(\partial_j\lambda_i-\partial_i\lambda_j\bigr)\,,\nonumber\\
a^{(1)\lambda}_{\phantom{(1)}i}&=&-\partial_i\eta\,,\nonumber\\
a^{(1)\eta}&=&0
\end{eqnarray}

\bigskip\noindent
and the corresponding matrix $f^{(1)}$ reads

\begin{equation}
f^{(1)}=\left(
\begin{array}{cccc}
0&\hskip-.5cm-\frac{1}{2}
\bigl(\delta_{ik}\delta_{jl}-\delta_{il}\delta_{jk}\bigr)
&\hskip-.5cm0&\hskip-.3cm0\\
{1\over2}\bigl(\delta_{ik}\delta_{jl}
-\delta_{ik}\delta_{jl}\bigr)&\hskip-.5cm0&\hskip-.5cm
{1\over2}\bigl(\delta_{jk}\partial_i
-\delta_{ik}\partial_j\bigr)&\hskip-.3cm0\\
0&\hskip-.5cm{1\over2}\bigl(\delta_{il}\partial_k
-\delta_{ik}\partial_l\bigr)
&\hskip-.5cm0&\hskip-.3cm\partial_i\\
0&\hskip-.5cm0&\hskip-.5cm\partial_k&\hskip-.3cm0
\end{array}
\right)
\,\delta^{(3)}(\vec x-\vec y\,)\,,
\end{equation}

\bigskip\noindent
where rows and columns follow the order $A$, $\pi$, $\lambda$,
$\eta$. This matrix is still singular. Let us consider a zero mode
like $\tilde v^{(1)}=(v_{kl},\,u_{kl},\,\omega_k,\,h)$. This will be
actually a zero mode if

\begin{eqnarray}
u_{ij}&=&0\,,\nonumber\\
v_{ij}&=&{1\over2}\bigl(\partial_j\omega_i
-\partial_i\omega_j\bigr)\,,\nonumber\\
\partial_ih&=&0\,.
\end{eqnarray}

\bigskip\noindent
Using the expression (\ref{6}) and the above conditions in order to
obtain possible new constraints, we get

\begin{eqnarray}
\int d^3\vec x\,\Bigl[v_{ij}(\vec x\,)\,
\frac{\delta}{\delta A_{ij}(\vec x\,)}
+\omega_i(\vec x\,)\frac{\delta}{\delta\lambda_i(\vec x\,)}
+\eta(\vec x\,)\frac{\delta}{\delta\eta(\vec x\,)}\Bigr]\,
\int d^3\vec y\,V^{(1)}&=&0\,,\nonumber\\
\Longrightarrow\hskip.5cm0&=&0\,.
\end{eqnarray}

\bigskip\noindent
As one sees, the zero modes do not lead to any new constraint. This
is a characteristic of gauge theories. So, in order to try to obtain
a nonsingular matrix, we fix the gauge. Let as then choose the
corresponding of the Coulomb gauge of the electromagnetic theory,
i.e.

\begin{equation}
\label{3.21}
\partial_iA^{ij}=0\,.
\end{equation}

\bigskip
Of course, since the Gauss' law constraints are not independent, the
gauge fixing conditions above cannot be independent either. We are
going to proceed as in the previous case, that is to say, we
introduce the constraints (\ref{3.21}) into the kinectic part of the
Lagrangian by means of Lagrange multipliers and restric these as in
(\ref{3.14}). The result is

\begin{equation}
{\cal L}^{(2)}=\bigl(\pi_{ij}-\partial_i\xi_j\bigr)\,\dot A^{ij}
-\dot\pi^{ij}\partial_i\lambda_j
-\dot\lambda^i\partial_i\eta
-\dot\xi^i\partial_i\rho-V^{(2)}\,,
\end{equation}

\bigskip\noindent
where

\begin{equation}
V^{(2)}=-{1\over2}\,\pi_{ij}\pi^{ij}
+{1\over2}\partial_iA_{jk}\,\partial^iA^{jk}\,.
\end{equation}

\bigskip\noindent
The term $\partial_iA_{jk}\partial^jA^{ik}$ of the previous Lagragian
was absorbed into the kinectic part of ${\cal L}^{(2)}$. Once more we
identify the new coefficients to calculate $f^{(2)}$. The final
result reads

\begin{eqnarray}
&&f^{(2)}=\nonumber\\
&&\nonumber\\
&&\left(
\begin{array}{cccccc}
0&\hskip-.5cm-{1\over2}\,
\bigl(\delta_{ik}\delta_{jl}-\delta_{il}\delta_{jk}\bigr)
&\hskip-.5cm0&\hskip-.5cm{1\over2}\,
\bigl(\delta_{jk}\partial_i-\delta_{ik}\partial_j\bigr)
&\hskip-.3cm0&\hskip-.3cm0\\
{1\over2}\,
\bigl(\delta_{ik}\delta_{jl}-\delta_{ik}\delta_{jl}\bigr)
&\hskip-.5cm0&\hskip-.5cm{1\over2}\,
\bigl(\delta_{jk}\partial_i-\delta_{ik}\partial_j\bigr)
&\hskip-.5cm0&\hskip-.3cm0&\hskip-.3cm0\\
0&\hskip-.5cm{1\over2}\,
\bigl(\delta_{il}\partial_k-\delta_{ik}\partial_l\bigr)
&00&\hskip-.5cm0&\hskip-.3cm\partial_i&\hskip-.3cm0\\
{1\over2}\,
\bigl(\delta_{il}\partial_k-\delta_{ik}\partial_l\bigr)
&\hskip-.5cm0&\hskip-.5cm0&\hskip-.5cm0
&\hskip-.3cm0&\hskip-.3cm\partial_i\\
0&\hskip-.5cm0&\hskip-.5cm\partial_k
&\hskip-.5cm0&\hskip-.3cm0&\hskip-.3cm0\\
0&\hskip-.5cm0&\hskip-.5cm0&\hskip-.5cm\partial_k
&\hskip-.3cm0&\hskip-.3cm0
\end{array}
\right)\nonumber\\
&&\nonumber\\
&&\hskip10cm\cdot\delta^{(3)}(\vec x-\vec y\,)
\end{eqnarray}

\bigskip\noindent
where rows and columns follow the order $A$, $\pi$, $\lambda$, $\xi$,
$\eta$, $\rho$. This matrix is not singular. It is the symplectic
tensor of the constrained theory. From its inverse we directly
identify the brackets

\begin{eqnarray}
\bigl\{A_{ij}(\vec x,t),\,\pi_{kl}(\vec y,t)\bigr\}
&=&\frac{1}{2}\Bigl[\Bigl(\delta_{ik}\delta_{jl}
-\delta_{il}\delta_{jk}\Bigr)
+\frac{1}{\nabla^2}\,\Bigl(\delta_{ik}\partial_j\partial_l
-\delta_{jk}\partial_i\partial_l\cr
&\phantom{=}&\phantom{{1\over2}\Bigl[\Bigl(\delta_{ik}\delta_{jl}}
-\delta_{il}\partial_j\partial_k
-\delta_{jl}\partial_i\partial_k\Bigr)\Bigr]\,
\delta^{(3)}(\vec x-\vec y\,)\,,\nonumber\\
\bigl\{A_{ij}(\vec x,t),\,\xi_k(\vec y,t)\bigr\}
&=&-\frac{1}{\nabla^2}\Bigl(\delta_{ik}\partial_j
-\delta_{jk}\partial_i\Bigr)\,
\delta^{(3)}(\vec x-\vec y\,)\,,\nonumber\\
\bigl\{\pi_{ij}(\vec x,t),\,\lambda_k(\vec y,t)\bigr\}
&=&\frac{1}{\nabla^2}\Bigl(\delta_{ik}\partial_j
-\delta_{jk}\partial_i\Bigr)\,
\delta^{(3)}(\vec x-\vec y\,)\,,\nonumber\\
\bigl\{\lambda_i(\vec x,t),\,\xi_j(\vec y,t)\bigr\}
&=&\frac{2}{\nabla^2}\Bigl(\delta_{ij}
+\frac{\partial_i\partial_j}{\nabla^2}\Bigr)\,
\delta^{(3)}(\vec x-\vec y\,)\,,\nonumber\\
\bigl\{\lambda_i(\vec x,t),\,\eta(\vec y,t)\bigr\}
&=&-\frac{\partial_i}{\nabla^2}\,
\delta^{(3)}(\vec x-\vec y\,)\,,\nonumber\\
\bigl\{\xi_i(\vec x,t),\,\rho(\vec y,t)\bigr\}
&=&-\frac{\partial_i}{\nabla^2}\,
\delta(\vec x-\vec y\,)\,.
\end{eqnarray}

\bigskip\noindent
The only really important bracket is the first one, that involves the
physical fields of the theory. It is a Dirac bracket in a sense that
it is strongly satisfied by the constraint relations. Since there is
no problem with ordering operators we can directly transform it to
commutator. Other brackets are not necessarily Dirac brackets. The
role of Lagrange multipliers are just to enlarge the configuration
space in order to make possible the definition of the symplectic
tensor, but they do not play any physical role in the theory.

\medskip
We mention that the first bracket above is the same one obtained in
\cite{Kaul} where the Dirac formalism was used.

\section{Conclusion}
We have considered the use of the symplectic formalism for reducible
systems. We have followed a similar procedure of hte BFV one where it
is necessary the introduction of {\it ghosts-of-ghosts}. Here we have
also introduced {\it Lagrange multipliers-of-Lagrange multipliers} in
order that the symplectic tensor could be defined. We have applied
the formalism to the antisymmetric tensor gauge field as an example.

\vskip 1cm
\noindent {\bf Acknowledgment:} We are in debt with R. Amorim and C.
Wotzasek for many useful comments. This work is supported in part by
Conselho Nacional de Desenvolvimento Cient\'{\i}fico e Tecnol\'ogico
- CNPq (Brazilian Research Agency).


\newpage
\begin{thebibliography}{30}
\bibitem{Dirac} P.A.M. Dirac, Can. J. Math. 2 (1950) 129; {\it
Lectures on quantum mechanics} (Yeshiva University, New York, 1964).
\bibitem{BFV} E.S. Fradkin and G.A. Vilkovisky, Phys. Lett. B55
(1975) 224; I.A. Batalin and G.A. Vilkovisky, Phys. Lett. B69 (1977)
309.
\bibitem{H} See also M Henneaux and C. Teitelboim, {\it Quantization
of gauge systems} (Princeton University Press, Princeton, 1992).
\bibitem{GSW} See, for example, M.B. Green, J.H. Schwarz and E. Witten,
{\it Superstring Theory} (Cambridge University Press, Cambridge,
1987) and references therein.
\bibitem{FJ} L.D. Faddeev and R. Jackiw, Phys. Rev. Lett. 60 (1988)
1692.
\bibitem{W} For a general review on this subject see, for example,
N. Woodhouse, {\it Geometric quantization} (Clarendon Press, Oxford,
1980).
\bibitem{CB} R. Floreanini and R. Jackiw, Phys. Rev. Lett. 59 (1987)
1873.
\bibitem{Gi} M.E.V. Costa and H.O. Girotti, Phys. Rev. Lett. 60
(1988) 1771.
\bibitem{Go} J. Govaerts, Int. J. Mod. Phys. A5 (1990) 3625.
\bibitem{Eu1} J. Barcelos-Neto and C. Wotzasek, Mod. Phys. Lett.
A7 (1992) 1737; Int. J. Mod. Phys. A7 (1992) 4981. See also  See also
H. Montani, Int. J. Mod. Phys. A8 (1993) 4319.
\bibitem{Outros} Further applications of the symplectic method with
constraints can be found in J. Barcelos-Neto and E.S. Cheb-Terrab, Z.
Phys. {\bf C54}, 133 (1992); M.M. Horta-Barreira and C. Wotzasek,
Phys. Rev. D45 (1992) 1410; C. Wotzasek, Phys. Rev. D45 (1992) 1410;
J. Barcelos-Neto, Phys. Rev. D 49 (1994) 1012.
\bibitem{Eu2} J. Barcelos-Neto and E. Vasquez, {\it Symplectic
quantization with higher derivatives}, Preprint IF/UFRJ. To appear in
Z. Phys. C.
\bibitem{Kaul} R.K. Kaul, Phys. Rev. D18 (1978) 1127.
\end{thebibliography}
\end{document}